\documentstyle[12pt]{article}

\begin{document}
\begin{titlepage}
\begin{center}
\rightline{CERN-TH-6838/93}
\vskip 2.cm

{\Large \bf
 q-Deformed  Brownian Motion}

\vskip 1.5cm
V. I. Man'ko\\
P. N. Lebedev Physical Institute,
Leninsky Prospect 53, 117924 Moscow, Russia
\\
\medskip
R. Vilela Mendes\footnote{
Permanent/Mailing address: CFMC, Av. Gama
Pinto 2, 1699 Lisboa Codex, Portugal}
\\
Theory Division, CERN, CH 1211 Geneva 23, Switzerland\\
\vskip 1.cm

\end{center}

\vskip 2.cm

\begin{abstract}
Brownian motion may be embedded in the Fock space of bosonic free
fields in one dimension. Extending this correspondence to a family
of creation and annihilation operators satisfying a  q-deformed
algebra, the notion of  q-deformation  is carried from the algebra
to the domain of stochastic processes. The properties of  q-deformed
Brownian motion, in particular its non-Gaussian nature and cumulant
structure, are established.

\end{abstract}
\vfill\leftline{CERN-TH-6838/93}
\leftline{March 1993}

\end{titlepage}
The concept of symmetry plays an essential role in the description
of physical phenomena. In most cases this symmetry is related to
covariance under the transformations induced by a Lie algebra. A
generalization of this mathematical structure, the  q-deformed
(or quantum) algebras, has recently emerged$^{[1-7]}$.  q-deformed
algebras, first discovered in the context of integrable lattice
models, were later identified as an underlying mathematical structure
in topological field theories$^{[8]}$ and rational conformal field
theories$^{[9]}$. Other attempts to apply the notion of  q-deformed
algebras cover a wide range of different domains, from space-time
symmetries$^{[10-12]}$ to gauge fields$^{[13]}$, to quantum chemistry
$^{[14]}$.
\\
In view of the actual and potential applications of q-deformation
in the context of Lie algebras and superalgebras, it is interesting
to ask whether the notion of q-deformation can also be extended
to other (non-algebraic) mathematical structures. In this paper
we try to extend this notion to stochastic processes. Our starting
point is the well-known embedding of Brownian motion in the Fock
space of bosonic free fields in one dimension$^{[15,16]}$. Extending
this correspondence to a time family of creation and annihilation
operators satisfying a  q-deformed  algebra we establish a  q-deformation
of Brownian motion.
\\
 q-deformed  creation and annihilation operators were defined by
several authors$^{[17-20]}$. They satisfy the algebra
$$a a^\dagger - q^{-1}  a^
\dagger a  = q^N \eqno (1.a)$$
$$a a^\dagger -  q  a^\dagger
a  =  q^{-N} \eqno (1.b)$$
where $N$ is the number operator
$$ [N, a^\dagger]
=  a^\dagger \hskip 2cm    [N,a] =- a\eqno (2)$$
The operators $a a^\dagger$ may be realized as infinite-dimensional
matrices on a vector space by $$
 a^\dagger \vert n> = \sqrt{[n+1]} \vert n+1>\hskip 2cm
a\vert n>
 = \sqrt{[n]}\vert n-1>\hskip 2cm     N \vert n>=n\vert n>
\eqno(3)$$
where we used the notation
$$[X]=X_q
 = {{ \sinh (X  \ln q)}
\over{ \sinh( \ln q)}}  \eqno (4)$$
$X$ being a number or an operator.  q-deformation  of single boson
operators is invariant under the replacement $q \rightarrow  q  ^{-1}$
and we write the algebra in an explicitly symmetric form which
will be useful later on.
$$a a^\dagger -   {{1}\over{2}}(q+q^{-1})
a^\dagger a   =   {{1}\over{2}}(q^N + q^{-N})
\eqno        (5)$$
(Notice that $(q+q^{-1})  =[2]$)
\\
We now consider a family \{ $a_\tau $ , $a^\dagger _\tau
$ \}  of  q-deformed  operators labelled by a continuous
time parameter and a scalar field
$$\phi_\tau
 =   a_\tau + a^\dagger _\tau \eqno  (6)$$
For the \{ $a_\tau $ , $a^\dagger _\tau $ \}
family we generalize the relations (5) and (2) to
$$a_{\tau _1}   a^\dagger _{\tau _2}   -
{{1}\over{2}}(q+q^{-1})   a^\dagger _{\tau _2}a_{\tau _1}   =
{{1}\over{2}}(q^N+q^{-N})   \delta (\tau _1-\tau _2)
\eqno (7)$$
$$[N,  a^\dagger _\tau ]   =
a^\dagger _\tau\hskip 2cm       [N,a_\tau ]=-a_\tau \eqno(8)
$$
This is the simplest extension of the relations to a family of  q-deformed
operators labelled by a continuous parameter. Other generalizations
of (5) are possible, involving for example braid relations at different
times. Notice also that, for our purposes of construction of a stochastic
process, no assumptions are needed concerning the commutation properties
of $a_{\tau _1}$ $a_{\tau _2}$ and $a^\dagger _{\tau _1}$ $a^\dagger
_{\tau _2}$ at different times.
\\
Smearing the fields with characteristic functions $\chi _{[0,t]}$
of the interval [0,t]
$$a_q(t)   =   a_q(\chi _{[0,t]})
 =   \int \limits_0\limits^td\tau  a_\tau   \eqno (9.a)$$

$$a_q^\dagger (t)   =   a_q^
\dagger (\chi _{[0,t]}) =\int \limits_0\limits^td\tau  a_
\tau ^\dagger   \eqno (9.b)$$
$$\phi _q(t)   =   \phi _q(\chi
_{[0,t]}) = \int \limits_0\limits^td\tau \phi _\tau
\eqno (10)$$
the algebraic relations become
$$
a_q(t_1) a_q^\dagger (t_2) - {{1}\over{2}}(q+q^{-1}) a_q^
\dagger (t_2)   a_q(t_1)   =   {{1}\over{2}}(q^N+q^{-N})   <\tau
_1\vert \tau _2> \eqno (11)$$
where $<\tau _1\vert \tau _2>$ $=$ min($t_1,t_2)$ .
\\
We now use (11) to construct a  q-deformation  of Brownian motion.
Let ($\Omega ,F_t,\mu ,B_t)$ be the usual Brownian motion. $\Omega
$ is the set of continuous functions vanishing at t$=0,$ $\mu $
is the Wiener measure and $F_t$ is the $\sigma$-ring  generated
by \{ $B_s:$ 0$\leq s\leq t\} .$ On the other
hand let (${\cal H},A_t,\psi _0,\phi _1(t))$ be the free quantum
field over K$=L^2([0,\infty ),{\bf  R}).$ $\phi _1(t)=\phi _1(\chi
_{[0,t]})$ (for q$=1),$ ${\cal H}$ is the symmetric Fock space over
K, $\psi _0$ is the Fock vacuum and $A_t$ is the $W^*-algebra$ generated
by \{ $\phi _1(s):0\leq s\leq t\} .$ Then$^{[16]}$,
interpreting $B_t$ as a multiplication operator in $L^2(\Omega ,F_t,
\mu )$, there is a unitary operator $V : L^2(\Omega ,F_t,\mu )\rightarrow
{\cal H}$ such that $VB_tV^{-1}=\phi (t)$. I.e. $\phi (t)$ as a
stochastic process with expectation
$$E(f( \phi (t)))
 = <\psi _0,f(\phi (t)) \psi _0> \eqno (12)$$
coincides with Brownian motion. For this identification of the free
scalar field with Brownian motion it is useful to characterize the
filtration $A_t$ by the conditional expectation of Wick products$^{[21]}$
$$E( : \phi _1(u_1)  ...  \phi _1(u_n):
\vert    A_t)   =  :  \phi _1(\chi _{[0,t]}u_1)  ...  \phi _1(\chi
_{[0,t]}u_n)  :\eqno     (13)$$

Recall that the Wick products span the algebra generated by $\phi
_1(u).$ Hence, by linearity, definition on Wick products suffices
to define conditional expectations on the complete algebra.
\\
We now use a minimal version of this correspondence to define q -deformed
Brownian motion.\newline
\underline{Definition}: q-deformed  Brownian motion
is the process ($
\Omega ,F_t,\mu _{\psi _0},\phi _q(t))$ where\newline
(i) $\phi _q(t)$ is the operator defined by (9$-11)$\newline
(ii) Expectations of field functionals $f(\phi _q)$ are obtained
by
$$
E(f( \phi _q))= <\psi _0,f( \phi _q) \psi _0>\eqno(14)$$
$\psi_0$
being defined by $a_q   \psi _0   =0$
\newline
(iii) The filtration $F_t$ is characterized by the conditional
expectations of Wick products
$$
E( : \phi _q(t_1)  ...  \phi _q(t_n):   \vert    F_s)   =  :  \phi
_q(\chi _{[0,s]}\chi _{[0,{t_1}]})  ...  \phi _q(\chi _{[0,s]}\chi
_{[0,{t_n}]})\eqno(15)$$
\centerline{
$--------------$}\\
Notice that the algebraic relations (11) allow all elements of the
algebra generated by $\phi _q(t)$ to be reduced to Wick products,
hence all conditional expectations may be computed. Notice also
that in this minimal definition the family $F_t$ of measurable events
is fixed in advance and we avoid an explicit realization of the
probability space $\Omega .$\newline
\underline{Theorem}\  \underline{1} :  q-deformed  Brownian
motion has : \newline
(i) zero mean, $E(\phi _q(t))=0$\newline
(ii) variance $E(\phi _q(t)$ $\phi _q(s))=$min$(s,t)$\newline
(iii) independent increments in the sense
\\ $
E(\{  \phi _q(t_1)-\phi _q(t_2)\} \{
\phi _q(t_3)-\phi _q(t_4)\} )= E (\phi _q(t_1)-\phi _q(t_2))
E( \phi _q(t_3)-\phi _q(t_4))=0$ \newline
if there is no overlap between the intervals [$t_1,t_2]$ and [$t_3,t_4]$
\\
(iv) is a martingale\newline
\\
Properties (i) to (iii) follow by a simple computation using (10)
(11) and (14). The martingale property is a consequence of (15).
If $s<t$
$$
E( \phi _q(t) \vert F_s) = \phi
_q(\chi _{[0,s]}\chi _{[0,t]}) = \phi _q(s)$$ \\
Theorem 1 summarizes the similarities of  q-deformed  Brownian motion
to the usual Brownian motion. The next result displays their main
differences as well as an explicit characterization in terms of
cumulants.\\
\eject
\underline{Theorem}\  \underline{2} :\newline
(i)  q-deformed  Brownian motion is not a Gaussian process\newline
(ii) The cumulants are
$$
E_T (\phi_q(t_1)...\phi_q(t_n))=\sum
c_{(i_1 i_2)...(i_{n-1} i_n)}
<t_{i-1}\vert t_{i_2}>...<t_{i_{n-1}}\vert t_{i_n}>
\eqno (16)
$$
where
$<t_i\vert  t_j>$=min$(t_i , t_j)$ , the sum is over all
different partitions of the set
($t_1...t_n$)
into pairs and the coefficients $c_{(i_1 i_2)...(i_{n-1} i_n)}$
are obtained by the following graphical rules\newline
(ii.a) For each term in (16) one draws the $<t_{i_k}\vert$ $t_{i_{k+1}}>$
contractions as follows\\

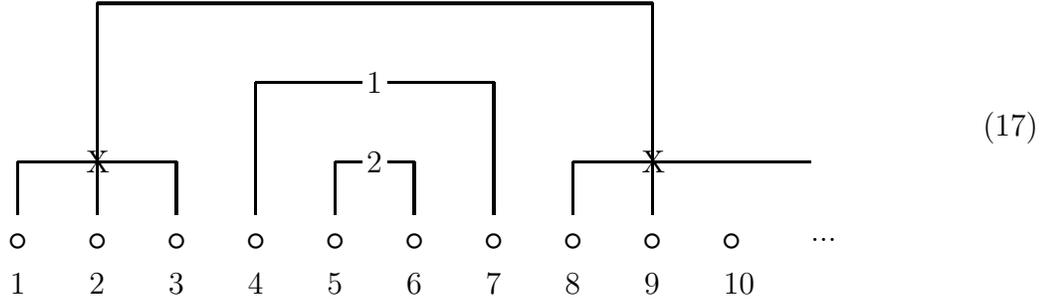
\begin{picture}(300,100)(-50,-15)
\thicklines
\put(0,0){\circle{5}}
\put(30,0){\circle{5}}
\put(60,0){\circle{5}}
\put(90,0){\circle{5}}
\put(120,0){\circle{5}}
\put(150,0){\circle{5}}
\put(180,0){\circle{5}}
\put(210,0){\circle{5}}
\put(240,0){\circle{5}}
\put(270,0){\circle{5}}
\put(300,0){...}
\put(365,40){(17)}
\put(26,26){X}
\put(236,26){X}
\put(132,26){2}
\put(132,56){1}
\put(-3,-20){1}
\put(27,-20){2}
\put(57,-20){3}
\put(87,-20){4}
\put(117,-20){5}
\put(147,-20){6}
\put(177,-20){7}
\put(207,-20){8}
\put(237,-20){9}
\put(267,-20){10}
\put(0,30){\line(1,0){60}}
\put(120,30){\line(1,0){10}}
\put(140,30){\line(1,0){10}}
\put(210,30){\line(1,0){90}}
\put(90,60){\line(1,0){40}}
\put(140,60){\line(1,0){40}}
\put(30,90){\line(1,0){210}}
\put(0,10){\line(0,1){20}}
\put(30,10){\line(0,1){80}}
\put(90,10){\line(0,1){50}}
\put(120,10){\line(0,1){20}}
\put(150,10){\line(0,1){20}}
\put(180,10){\line(0,1){50}}
\put(240,10){\line(0,1){80}}
\put(210,10){\line(0,1){20}}
\put(60,10){\line(0,1){20}}
\end{picture}
\\
\vskip .5cm
(ii.b) For each crossing of lines there is a factor
$\{ {{1}\over{2}}(q+q^{-1})-1\}$ in
$c_{(i_1 i_2)...(i_{n-1} i_n)}$\\
(ii.c) For each contraction contained at depth $\alpha $ inside
other contractions there is a factor
$\{ {{1}\over{2}}(q^\alpha +q^{-\alpha }) - 1\}$
 in $c_{(i_1  i_2 )...(i_{n-1} i_n)}$ \newline
(ii.d) If there are no crossings nor inner contractions the coefficient
$c_{(i_1 i_2)...(i_{n-1} i_n)}$ vanishes.\newline
\centerline{
$--------$}\newline
In the example (17) the contribution of the diagram to the coefficient
is
$$
\{ {{1}\over{2}}(q+q^{-1})-1\}^2\{
{{1}\over{2}}(q^2+q^{-2})-1\}\{ {{1}\over{2}}
(q+q^{-1})-1\}$$
the first factor coming from the crossings and the last two from
the inner contractions at level 2 and 1.
A necessary and sufficient condition for a process to be Gausssian
is that it possesses cumulants of all orders and that they vanish
for order higher than 2. Computing the 4-time correlation one obtains
using (11) and (14)
$$
E(t_1 t_2 t_3 t_4)=E(\phi_q(t_1) \phi_q(t_2) \phi_q(t_3)
 \phi _q(t_4)) =$$
$$<t_1\vert t_2><t_3\vert t_4>+{{1}\over{2}}(q+q^{-1})<t_1
\vert t_3><t_2\vert t_4>+{{1}\over{2}}(q+q^{-1})<t_1\vert t_4><t_2
\vert t_3>$$
implying that the cumulant
$$
 E_T(t_1t_2t_3t_4)=E(t_1t_2t_3t_4)-E(t_1t_2)E(t_3t_4)-
 E(t_1t_3)E(t_2t_4)-E(t_1t_4)E(t_2t_3)$$
does not vanish. Hence the process is not Gaussian.
\\
The explicit expression for the cumulants of arbitrary orders is
obtained by systematic reduction of the expectation values using
the algebraic relations (11). The ``crossing lines'' factor comes
from the coefficient of the second term in the l.h.s. of (11) and
the ``inner contractions'' factor from the r.h.s. together with
(8).
\\
As a final remark we point out that using  q-fermions $b$ and $b^
\dagger $ with
$$
b b^\dagger+q  b^\dagger b = q^M\hskip 2cm [M,b^\dagger]
=b^\dagger\hskip 2cm    [M,b] =-b  \eqno (18)$$
and a generalization along the lines of (6 -11)  one may construct
a  q-deformation  of the non-commutative Clifford process$^{[21]}$.
We leave the details to the interested reader.

One of the authors (V.I.M.) thanks Centro de Fisica da Materia
Condensada (Lisboa) for hospitality.\\

\medskip
{\Large\bf References}\\
\medskip
[1] E. K. Sklyanin, Funct. Anal. Appl. 16 (1982) 263.\newline
[2] P. P. Kulish and N. Yu. Reshetikhin, J. Sov. Math. 23 (1983)
2435.\newline
[3] V. G. Drinfel'd, Sov. Math. Dokl. 32 (1985) 254; 36 (1988) 212;
J. Sov. Math 41 (1988) 898.\newline
[4] M. Jimbo, Lett. Math Phys. 10 (1985) 63; 11 (1986) 247.\newline
[5] S. L. Woronowicz, Comm. Math. Phys. 111 (1987) 613.\newline
[6] L. D. Faddeev, N. Yu. Reshetikhin and L. A. Takhtajan, Alg.
Anal. 1 (1988) 129. \newline
[7] Yu. I. Manin, Ann Inst. Fourier 37 (1987) 191; Comm. Math. Phys.
123 (1989) 163.\newline
[8] E. Witten, Comm. Math. Phys. 121 (1989) 351.\newline
[9] L. D. Faddeev, Comm. Math. Phys. 132 (1990) 131 and references
therein.\newline
[10] P. Podles and S. L. Woronowicz, Commun. Math. Phys. 130 (1990)
381.\newline
[11] C. Gomez and G. Sierra, Phys. Lett. B255 (1991) 51.\newline
[12] J. Lukierski, A. Nowicki, H. Ruegg and V. N. Tolstoy, Phys.
Lett. B264 (1991) 331.\newline
[13] I. Ya. Aref'eva and I. V. Volovich, Mod. Phys. Lett A6 (1991)
893; Phys. Lett. 264B (1991) 62.\newline
[14] M. Kibler and T. N\'egadi, Lyon preprint LYCEN 9121 (1991).
\newline
[15] T. Hida, ``Brownian motion'', Springer, Berlin 1980.\newline
[16] R. F. Streater, Acta Physica Austriaca, Suppl. XXVI (1984)
53.\newline
[17] A. J. Macfarlane, J. Phys. A: Math. Gen. 22 (1989) 4581.\newline
[18] L. C. Biedenharn, J. Phys. A: Math. Gen. 22 (1989) L873.\newline
[19] C.-P. Sun and H.-C. Fu, J. Phys. A: Math. Gen. 22 (1989) L983.
\newline
[20] P. P. Kulish and N. Yu. Reshetikhin, Lett. Math. Phys. 18 (1989)
143.\newline
[21] R. F. Streater, ``The fermion stochastic calculus I'' in ``Stochastic
Processes $-$ Mathematics and Physics'', S. Albeverio, Ph. Blanchard
and L. Streit (Eds.), Springer Lecture Notes in Mathematics 1158,
Berlin, 1986. \newline

\end{document}